\documentclass[sigplan,nonacm,authorversion]{acmart}
 
\usepackage{tikz}
\usepackage{algorithm}
\usepackage{algpseudocode}
\usepackage{mathtools}
\usepackage{graphicx}
\usepackage{multirow}
\usepackage{ulem}

\usetikzlibrary{backgrounds,fit,calc,positioning,arrows.meta}

\author{Srinidhi N}
\affiliation{
    \institution{Advanced Micro Devices, Inc.}
    \country{India}
}
\email{Srinidhi.N@amd.com}

\author{Sunanda Somwase}
\affiliation{
    \institution{Advanced Micro Devices, Inc.}
    \country{India}
}
\email{SomwaseSunanda.Vishnu@amd.com}

\author{Hemant Kumar Chandrakant Chandavar}
\affiliation{
    \institution{Advanced Micro Devices, Inc.}
    \country{India}
}
\email{hkcc375@gmail.com}

\author{Eashan Dash}
\affiliation{
    \institution{Advanced Micro Devices, Inc.}
    \country{India}
}
\email{Eashan.Dash@amd.com}

\author{Varun Agrawal}
\affiliation{
    \institution{Advanced Micro Devices, Inc.}
    \country{India}
}
\email{Varun.Agrawal@amd.com}

\author{Prakash Raghavendra}
\affiliation{
    \institution{Advanced Micro Devices, Inc.}
    \country{India}
}
\email{Prakash.Raghavendra@amd.com}

\definecolor{loopccd}{HTML}{1B6FB3}
\definecolor{loopmacro}{HTML}{B3541B}
\definecolor{loopmicro}{HTML}{2E8B57}

\newcommand{\codename}{\textsc{Waypoint}}
\newcommand{\codenamebf}{\textbf{\codename}}
\newcommand{\codenamesimple}{\codename{}-\textsc{Strict}}
\newcommand{\codenamerefined}{\codename{}-\textsc{Tile}}

\newcommand{\libraryname}{\textsc{OPAM}}

\setcopyright{cc}
\makeatletter
\renewcommand{\@copyrightpermission}{%
    \textcopyright{} 2026 Advanced Micro Devices, Inc. All rights reserved.
    AMD, the AMD Arrow logo, EPYC, AOC-BLAS, AOC-DLP, PACE, and combinations
    thereof are trademarks of Advanced Micro Devices, Inc. Other product names
    used in this publication are for identification purposes only and may be
    trademarks of their respective companies.}
\makeatother

\begin{document}
\title{\codenamebf{}: Analytically Modeling GEMM for CPUs}

\begin{abstract}

  High-performance General Matrix Multiplication (GEMM) implementations rely on breaking down GEMM into carefully selected tiles to achieve peak performance. We present \codename{}, a new analytical model to automatically identify the optimal tile size for a given hardware and input matrix shape. \codename{} estimates the achievable performance for a given tile choice by capturing all the architectural effects including interactions with memory hierarchy, instruction-level parallelism, and multi-threading. Using the estimates as a proxy, we search through the space of possible tile sizes to identify the one that maximizes performance. Additionally, \codename{} helps hardware designers estimate realistic performance on experimental hardware optimizations. We build a new GEMM library \libraryname{} for CPUs that uses \codename{} to automatically break down GEMM and achieve performance that matches state-of-the-art GEMM libraries and helps accelerate LLM inference on CPUs achieving $1.36 - 1.98 \times$ speedups.
\end{abstract}

\maketitle %

\section{Introduction}
\label{sec:intro}

Matrix Multiplication operations make up for a large part of the core computation in several modern machine learning based applications such as image processing, recommendation systems, and more recently, Large Language Model (LLM) inference. To ensure that these applications run efficiently, it is crucial to optimize the performance of General Matrix Multiplication (GEMM) operations on the underlying hardware.
While the industry has moved towards developing specialized hardware accelerators such as GPUs,
CPUs remain important for cost-sensitive use cases and small-scale inference workloads.
Hardware vendors continue to enhance CPU architectures with new features aimed at improving GEMM performance.

On CPUs, improving GEMM performance involves careful coordination between the hardware and the software that implements GEMM operations. Hardware designers continue to innovate and introduce features that push the peak performance achievable. At the same time, software implementations optimize the use of these hardware features to achieve high performance.

A naive 3-loop implementation of the matrix multiplication algorithm is far from achieving peak performance. Instead, the libraries use specialized hardware features (e.g. vector/matrix instructions), and carefully organize the input matrices in memory so as to reduce the cost of accessing them for the computation. They do so by breaking down the GEMM operation into smaller units called tiles. For each tile, the libraries use vector/matrix instructions to compute the output, and between successive tiles, they ensure that memory accesses are efficient.

The tile sizes are critical to achieving high performance and need to be chosen carefully. Selecting suboptimal tile sizes can lead to inefficient memory accesses and under-utilization of the hardware's computational capabilities. The right tile depends both on the hardware and the specific input matrix size that the application invokes~\cite{analytical_model_2}. Therefore, an ideal solution would adapt the tile sizes not only to the hardware but also to specific input sizes. %

Existing state-of-the-art GEMM libraries~\cite{blis,goto-gemm} identify the tile sizes that are either optimal for a given hardware, or the specific class of shapes (e.g., tall and skinny~\cite{tallandskinny}, irregular~\cite{analytical_model_2}) but do not adapt them to the application's needs. Many techniques are used to identify the tile sizes for a given hardware such as - manual expert-driven tuning resulting in handwritten assembly, automated empirical analysis~\cite{autotuning,atlas}, and analytical modeling~\cite{analyticalmodellingisenough,analytical_model_1,analytical_model_2}.

In this paper we propose a new analytical model to derive optimal tile sizes that are tailored not just for the hardware but also for the specific input matrix sizes that the application uses. Our model addresses another crucial challenge that the hardware designers face when innovating to accelerate performance on future hardware. They need to be able to estimate the realistic performance achievable on experimental hardware designs. They routinely propose new ISA extensions (e.g. AMX~\cite{amx}, ACE~\cite{ace_spec}, APX~\cite{apx}) aimed at optimizing GEMM performance.

Existing analytical models address the challenge partially. They provide a Roofline~\cite{roofline} estimate of the maximum achievable performance. But depending on the matrix sizes, the realistic achievable performance is far from the Roofline.

Building a realistic performance model requires accounting for different architectural effects that are accompanied with a GEMM operation such as cache effects, pipelining, multi-threaded execution, and memory bandwidth limitations.
The traditional Roofline model makes simplistic assumptions to over-approximate the peak performance achievable. It assumes that either the cost of computation dominates the cost of accessing data (i.e. compute-bound) or vice versa (i.e. memory-bound), which rarely holds true in practice. Therefore, a more nuanced model that captures both computation and memory access costs accurately is necessary to predict realistic GEMM performance.

In what follows, we describe our contribution - a new analytical model \codenamebf{} that is parametric both in the hardware architecture, and the application's needs (input matrix sizes). \codename{} provides accurate estimates of the realistic performance achievable, aiding hardware designers, and identifies the optimal tile sizes, enabling automatic and efficient GEMM implementations.

We build our model in two stages. First, we develop \codenamesimple{}, to provide realistic estimates of GEMM performance on a given hardware independent of the software implementation. We prove theoretically that \codenamesimple{} is tighter than the Roofline model and show empirically that state-of-the-art GEMM libraries realize these estimates. In compute-bound scenario, \codenamesimple{} accounts for the cost of additional instructions that are always issued with the main compute instruction. Similarly for the memory-bound scenario, \codenamesimple{} uses more accurate bandwidth values to calculate the time it takes to fetch data. As a result, in both scenarios, \codename{} provides estimates that are tighter than the Roofline.

Second, we refine \codenamesimple{} for a tile based GEMM implementation resulting in \codenamerefined{} that estimates GEMM performance for a specific tile size. Using \codenamerefined{}'s estimates, we search through possible tile sizes to select the optimal one. Building \codenamerefined{} is challenging because it requires accurately modeling the interaction between the tile size and the memory hierarchy, including cache levels, cache line sizes, and memory bandwidth limitations. \codenamerefined{} accurately estimates the number of cache misses that are imposed by a tile size and adds the associated miss costs. In building \codenamerefined{}, we identify several insights that allow us to rule out certain tile sizes and narrow down the search for an optimal one.

Empirically, we evaluate the accuracy of both \codenamesimple{} and \codenamerefined{} against state-of-the-art GEMM libraries. We find that \codenamesimple{} estimates are consistently closer to the observed performance (a geomean of $1.096\times$). Similarly, we compare the estimate of the best tile size identified by \codenamerefined{} and find that the model is accurate in predicting realistic performance (a geomean of $1.012\times$). These accuracy results allow us to confidently use the models for identifying the optimal tile size and estimating the performance on future hardware.

We develop a GEMM library \libraryname{} that uses \codenamerefined{} to select the best tile size for each input matrix size and hardware configuration. \libraryname{} performs comparable to state-of-the-art GEMM libraries across different hardware configurations and input sizes. Further on new hardware configurations, \libraryname{} is able to quickly adapt and achieve near-peak performance without any manual tuning. Finally, we use \libraryname{} to accelerate LLM inference on CPUs achieving $1.36 - 1.98 \times$ speedups on new hardware.

To summarize, our contributions are,
\begin{itemize}
    \item An accurate implementation agnostic analytical model \codenamesimple{}, to estimate GEMM performance that is parametric in both the hardware architecture and the application's needs. A more refined model, \codenamerefined{} that is specific to a tile-based GEMM implementation and used in practice to identify the optimal tile-size.
    \item A GEMM library \libraryname{} that uses \codenamerefined{} to automatically select optimal tiling parameters for a given GEMM shape and hardware configuration, achieving near-peak performance without extensive manual or automated tuning.
    \item Using \libraryname{}, we accelerate LLM inference on current and future CPUs.
\end{itemize}

\section{Analytical Model}
\label{sec:model}

The Roofline model~\cite{roofline} distinguishes a computation between a compute-bound scenario, where performance is limited by the processing capacity of the CPU, and a memory-bound scenario, where performance is limited by the memory bandwidth. It provides upper bounds in either case based on the peak computational throughput and memory bandwidth of the hardware.

Building on the Roofline model specifically for GEMM computations, we first develop an implementation-agnostic \codenamesimple{} that refines the cost in both compute-bound and memory-bound scenarios. Then, we further refine the compute cost assuming a tiled GEMM implementation to obtain a \codenamerefined{} that captures accurately the number of cache misses and the associated costs. We prove theoretically that the \codenamesimple{} is tighter than the Roofline model and empirically show (in Section~\ref{sec:eval}) that the \codenamerefined{} is accurate.

\subsection{\codenamesimple{}}

The \codenamesimple{} is parametric in the hardware configuration, including cache sizes, memory bandwidth, and compute instructions and provides an estimate of the number of cycles a GEMM operation will take on the given hardware.

Formally a hardware configuration $H$ consists of the following constants:
\begin{itemize}
    \item The instruction reciprocal throughput for the core compute instruction $\tau_{fma}$ and the dimensions of the instruction $\delta_{fma}$. The instruction is defined for each data-type and instruction set architecture (ISA) of the CPU. For example, in an x86-ISA based CPU using AVX512 family of SIMD instructions, the instruction reciprocal throughput for a single precision FMA instruction in AMD `Zen 5' architecture is 0.5 cycles/instruction and contains $16\times 1$ elements.
    \item Register file information such as the number of registers available to issue FMA instructions $N_{reg}$ and width of the registers in bits $W_{reg}$.
    \item Cache sizes ($S_{L1}, S_{L2}, S_{L3}$) and access reciprocal throughput ($T_{L1}, T_{L2}, T_{L3}$) for the different levels of the memory hierarchy. For example, on AMD `Zen 5' cores, the L1 data cache is 48 KB with an access reciprocal throughput of 0.5 cycles.
    \item Memory bandwidth information, including the peak bandwidth per memory channel $D_{bw}$ and the number of channels $D_n$.
    \item Maximum number of cores $C_{max}$ and the frequency $f$ of the CPU. The frequency is used to convert the number of cycles to time.
\end{itemize}

We define a GEMM operation $op$ as one that computes an output matrix $C = a(A \cdot B) + bC$ of size ${M \cdot N}$ where the input matrices are of size ${M \cdot K}$ and ${K \cdot N}$ using $T$ threads on the data-type $d$ of width $W_d$ bits, and scalars $a$ and $b$. We assume for simplicity that $a = 1$ and $b = 0$.

Given the hardware and GEMM operation, the \codenamesimple{} computes the latency $L_{op}$ of the GEMM operation as follows:

\[L(op) = \max\left(\frac{L_c(op)}{op.T}, L_m(op)\right) + L_{mt}(op)\]

where $L_c$ is the compute latency, $L_m$ is the memory latency, and $L_{mt}$ is the threading overhead for the GEMM operation $op$. In what follows, we describe each of the components in detail.

\subsubsection{Compute latency}

The \codenamesimple{} assumes a deterministic compute instruction to be used for a given data-type and hardware architecture e.g. AVX512 VNNI is used for Int8 GEMM on x86-ISA based CPUs. The exact mapping can be provided as input. We first compute the number of compute instructions $N_{fma}$ required to perform the GEMM:

\[ N_{fma} = M \cdot N \cdot K / \delta_{fma}\]

Where $\delta_{fma}$ is used to denote the number of elements in an instruction e.g. In an AVX512 VNNI instruction that computes a column-wise dot product of two $4 \times 16$ matrices of 8-bit integers, $\delta_{fma} = 4 \times 16 = 64$.

\paragraph{Instruction Modularity} By allowing the user to define the dimensions of the core compute instruction, we accommodate new advances in the x86-ISA architecture that enable matrix instructions such as AMX, ACE, Arm SME. A matrix instruction can compute a matrix product of two matrices of size $m_{fma} \cdot k_{fma}$ and $k_{fma} \cdot n_{fma}$ in a single instruction. Therefore, the number of compute instructions is $N_c = M \cdot N \cdot K / (m_{fma} \cdot k_{fma} \cdot n_{fma})$.

The latency of the compute instructions is computed as $N_{fma} * \tau_{fma}$ where $\tau_{fma}$ is the instruction reciprocal throughput for the core computation.

In addition to the latency of the compute instruction, the total compute latency $L_c$ includes the latency of support instructions for zeroing the registers $L_z$, loading the input matrices $L_{din}$ into the registers (assuming perfect cache behavior), and storing the output matrix back to memory $L_{dout}$. We assume that stores hit the nearest cache and thus incur the lowest latency.

\subsubsection{Memory latency}

The memory latency $L_m$ is computed using the peak memory bandwidth applicable for the GEMM operation.

\[L_m(op) = \frac{(M \cdot K + K \cdot N + M \cdot N) \cdot W_d \cdot f}{D_{bw} \cdot d_n}\]

where $d_n \leq D_n$ is the number of memory channels accessible to the GEMM operation.

When running multithreaded GEMM programs, the bandwidth available to each thread may be lower than the peak bandwidth because of contention among threads accessing memory. We capture the effect of this contention using a model that distributes work across different memory channels, allowing us to predict data-access latency more accurately.

For example, on AMD `Zen 5' processors, if a GEMM operation uses fewer than eight threads (the size of a single Core Complex Die, or CCD), it can utilize the memory bandwidth of a single CCD. If more threads are used, the operation may need to access multiple CCDs, potentially increasing effective memory latency because of contention across memory channels.

\subsubsection{Threading overhead}

To compute the threading overhead, we assume that there is no data contention between threads. This assumption is grounded in practice, as it reflects the behavior of well-optimized GEMM implementations. When the assumption does not hold, the threads must synchronize to read and write data, incurring additional overhead.

It follows from our assumption that each thread consumes the entire $K$ dimension to compute the result and the input matrices are split only in the $M$ and $N$ dimensions. Let $M_{t}$ and $N_{t}$ be the splits across the $M$ and $N$ dimensions of the output matrix for each thread.

The data needed by each thread is $S_{A_{t}} = M_{t} \cdot K \cdot W_d$ and $S_{B_{t}} = K \cdot N_{t} \cdot W_d$. The data that does not fit in the caches, $\Delta = S_{A_{t}} + S_{B_{t}} - (S_{L3} + S_{L2} + S_{L1})$, is repeatedly accessed from memory by multiple threads, incurring significant overhead. We define the threading overhead as follows:

\[L_{mt} = \min(\Delta \cdot M_{t}, \Delta \cdot N_{t}) \cdot \frac{1}{D_{bw} \cdot d_n}\]

To capture the scenario in which the $K$ dimension is also split, we account for the cost of the reduction operation that follows the partial computations across different threads. This reduction typically involves summing the partial results along the $K$ dimension, which introduces additional memory accesses and synchronization overhead.

\subsubsection{Theoretical results}

The Roofline model estimates peak performance for a given GEMM operation as,

\[P = \min(C_{peak}, B_{peak} \cdot I(op)) = \frac{2 \cdot M \cdot N \cdot K \cdot f}{L^r(op)}\]

where $C_{peak}$ is the peak compute performance of the CPUs, $B_{peak}$ is the peak memory bandwidth of the CPU, $I(op)$ is the operational intensity of the GEMM operation, and $L^r(op)$ is the Roofline latency. Formally,
\[C_{peak} = \frac{2\cdot \delta_{fma} \cdot op.T \cdot f}{\tau_{fma}} \quad B_{peak} = \max_D(D_{bw} \cdot D_n)\]

\[I(op) = \frac{2 \cdot M \cdot N \cdot K}{(M \cdot K + K \cdot N + M \cdot N) \cdot W_d}\]

Substituting the definitions, the resulting latency estimate $L^r(op)$ of the Roofline model is:

\[
    L^r(op) =
    \begin{dcases}
        \frac{M\cdot N \cdot K \cdot \tau_{fma}}{\delta_{fma} \cdot T}                           & \text{if } I > \frac{C_{peak}}{B_{peak}} \\
        \frac{(M\cdot N + M \cdot K + K\cdot N) \cdot W_d \cdot f}{D^{max}_{bw} \cdot D^{max}_n} & \text{otherwise}
    \end{dcases}
\]

To compare, the \codenamesimple{} is designed to provide more realistic estimates of latency in comparison to the Roofline model. We capture the result in the following theorem,

\begin{theorem}
    For any GEMM operation $op$ on a given hardware configuration $H$, $L^r(op) \leq L(op)$.
\end{theorem}

The theorem follows trivially from the definitions of $L(op)$ and $L^r(op)$. In the compute bound case, the \codenamesimple{} accounts for additional cost in the compute latency $L_c$ such as the cost of zeroing registers, loading, and storing data. Similarly, in the memory bound case, the effective bandwidth is at most $D_{bw} \cdot D_n$. In both cases, the additional threading overhead $L_{mt}$ might be non-zero. Therefore, the \codenamesimple{} is tighter than Roofline.

\subsubsection{Soundness of the assumptions}

\begin{figure}
    \includegraphics[width=\linewidth]{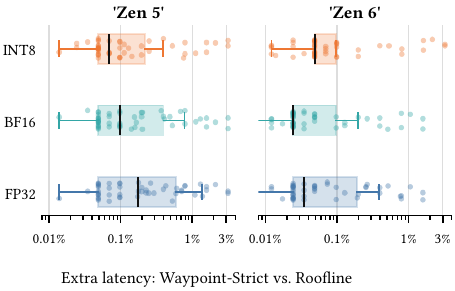}
    \caption{Additional latency captured by the \codenamesimple{} relative to Roofline across GEMM shapes, separated by hardware and data type. Boxes show the interquartile range and median; points show individual shapes.}
    \label{fig:roofline-ratio}
\end{figure}

Figure~\ref{fig:roofline-ratio} reinforces the theoretical result empirically. We compare the latency estimates of the \codenamesimple{} and the Roofline model on two different hardware configurations for many different shapes. The modest improvement over Roofline (upto $3\%$) is significant since performance of optimized GEMM libraries match the estimates (with $0.03\%$ gap) for specific input matrix sizes.

\subsection{\codenamerefined{}}

While the \codenamesimple{} provides a tighter-than-roofline estimate, it does not provide a mechanism to realize the estimates. Therefore, we refine the \codenamesimple{} to account for a tiling-based GEMM implementation.
The \codenamerefined{} helps identify the best tiling parameters for a given GEMM operation on a given hardware configuration. Using the tiling parameters, we construct a GEMM implementation (Section~\ref{sec:library}) that demonstrates the realizability of the estimates.
To identify the best tiling parameters, we augment the GEMM operation with three levels of tiling parameters: the register tiles $(M_r, N_r)$ that form the two inner loops of Algorithm~\ref{alg:gemm-loop}, the cache-tile parameters $(M_c, N_c, K_c)$, and the multithreading-tile parameters $(M_t, N_t)$. The \codenamerefined{} computes the latency of the GEMM operation for a given tiling option. Note that Algorithm~\ref{alg:gemm-loop} is one possible instantiation of the tiled implementation. In our implementation, we select the loop ordering that works best for a given tile parameter set.

Given the \codenamerefined{}, we then search the space of possible tiling options to identify the best. While theoretically, the search space suffers from combinatorial explosion, we describe heuristics to prune the search. In Section~\ref{sec:eval}, we also show that the best tiling parameters identified using the \codenamerefined{} improve over the state-of-the-art GEMM libraries across many different matrix shapes and hardware configurations. \codenamerefined{} makes certain assumptions of the behavior of the underlying hardware and implementation such as strided data accesses, fixed frequency workload, and using the best compute instruction available (e.g. AVX512 on x86 Systems). For the sake of brevity, we defer the details to Appendix~\ref{sec:appendix_c}

\begin{algorithm}
    \caption{Tiled GEMM: $C \mathrel{+}= A \cdot B$}
    \label{alg:gemm-loop}
    \algrenewcommand\algorithmicdo{}
    \begin{algorithmic}[1]
        \Require $A \in \mathbb{R}^{M \times K}$, $B \in \mathbb{R}^{K \times N}$, $C \in \mathbb{R}^{M \times N}$
        \Require \textcolor{loopccd}{Thread tile $(M_t, N_t )$}, \textcolor{loopmacro}{Macro tile $(M_c, N_c, K_c)$}, \textcolor{loopmicro}{Micro tile $(M_r, N_r, K_r)$}
        \For{\textcolor{loopccd}{$j_d \gets 0$ \textbf{to} $N - 1$ \textbf{step} $N_t $}}
        \For{\textcolor{loopccd}{$i_d \gets 0$ \textbf{to} $M - 1$ \textbf{step} $M_t$}}
        \For{\textcolor{loopmacro}{$j_c \gets j_d$ \textbf{to} $j_d + N_t  - 1$ \textbf{step} $N_c$}}
        \For{\textcolor{loopmacro}{$i_c \gets i_d$ \textbf{to} $i_d + M_t - 1$ \textbf{step} $M_c$}}
        \For{\textcolor{loopmacro}{$k_c \gets 0$ \textbf{to} $K - 1$ \textbf{step} $K_c$}}
        \For{\textcolor{loopmicro}{$j_r \gets j_c$ \textbf{to} $j_c + N_c - 1$ \textbf{step} $N_r$}}
        \For{\textcolor{loopmicro}{$i_r \gets i_c$ \textbf{to} $i_c + M_c - 1$ \textbf{step} $M_r$}}
        \State $C_{[i_r,\,j_r]} \mathrel{+}= A_{[i_r,\,k_c]} \cdot B_{[k_c,\,j_r]}$
        \EndFor
        \EndFor
        \EndFor
        \EndFor
        \EndFor
        \EndFor
        \EndFor
    \end{algorithmic}
\end{algorithm}

Apart from providing accurate estimates of GEMM operation for a given tiling option, the \codenamerefined{} also ranks the different tiling options accurately, based on their expected performance. We show empirically that the estimates of the best-tiling options are accurate within $5\%$ for existing hardware configurations.

The \codenamerefined{} reformulates the compute cost $L_c$ taking the tiling parameters into account. Specifically, we first break up the computation into micro-kernels -- the inner most loop that iterates over the $(M_r, N_r, K_c)$ dimensions. Then, for each micro-kernel, we account for additional costs such as cache access overheads when the tiling parameters are not optimal. Further, we add overheads for data movement costs at the macro-kernel level -- the outer $(M_c, N_c)$ loops. We redefine the compute cost $L_c$ as follows,
\[L_c = L_{\mu} * N_{\mu} + L_{\omega} * N_{\omega}\]
where $L_{\mu}$ and $N_{\mu}$ are the latency and number of micro-kernels, and $L_{\omega}$ and $N_{\omega}$ are the latency and number of macro-kernels.

We identify several analytical insights that capture the architectural behavior of the CPU and help prune the search space for the best tiling parameters. We describe the \codenamerefined{} in detail below.

\subsubsection{Micro-kernel cost}

The micro-kernel loop computes an output matrix of size ${M_r \times N_r}$ from input matrices of sizes ${M_r \times K_c}$ and ${K_c \times N_r}$. The micro-kernel is the innermost loop of the GEMM operation, where the bulk of the computation occurs, and is typically implemented in handwritten assembly code optimized for the specific hardware. The data needed for the micro-kernel is fetched from the L1 cache (closest to the CPU).

The handwritten assembly code for the micro-kernel has the following segments: (1) zero/load instructions to initialize the registers used to store the result; (2) prefetch instructions to fetch future data into the cache; (3) load instructions to load the input matrices into registers; (4) FMA instructions to perform the core computation; and (5) store instructions to store the result back to memory.

We associate a cost with each of the above segments and compute their sum as the total cost of the micro-kernel. All components depend on the number of registers used in the hand-written assembly in relation to the number of registers available in the hardware.

\paragraph{Reuse Overhead}
A key component of the micro-kernel latency cost is accurately modeling the cost of loading data. Optimal kernels under a perfect cache incur no cost to load data because of out-of-order execution. However, depending on the size of the data loaded (in terms of cache lines), a fraction of the load instructions will miss the cache and incur additional costs. We refer to this cost as reuse overhead. The reuse overhead is computed as the number of cache lines that incur a cache miss multiplied by the effective access latency from the L2 cache.

Let $A_{reg}$ and $B_{reg}$ to be the number of registers used to store data from the input matrices $A$ and $B$ (derived using $M_r$ and $N_r$ respectively). Then, then number of registers used to store the result, $C_{reg}$ is $C_{reg} = A_{reg} \cdot B_{reg}$. Together, the total number of registers used in one loop iteration of the micro-kernel is $R_{used} = A_{reg} + B_{reg} + C_{reg}$.

Two equations govern the computation of reuse-overhead and, in effect, determine the optimal values of $M_r$ and $N_r$. First, a lower bound on the register pressure, defined as the ratio of input and output registers. Second, an upper bound on the load-to-fma latency ratio, defined as the ratio of the load instruction latency to the FMA instruction latency. Formally,

\begin{lemma}{\textbf{Register pressure equation:}} The ratio of input and output registers is lower bounded by an architecture-dependent constant $\alpha$.
    \[\frac{A_{reg} + B_{reg}}{C_{reg}} \geq \alpha\]
\end{lemma}
For a given hardware architecture $H$, we calculate the bound $\alpha$ based on integer solutions to the equation,
\[A_{reg} + B_{reg} + C_{reg} \leq N_{reg}\]

\begin{lemma}{\textbf{Load-to-fma latency ratio:}} The ratio of the load instruction latency to the FMA instruction latency is upper bounded by 1.
    \[\frac{(A_{reg} + B_{reg})\cdot \tau_{load}}{C_{reg} \cdot \tau_{fma}} \leq 1\]
\end{lemma}

When either of the conditions are violated, the micro-kernel stalls to load the data, incurring a non-zero reuse cost. When register pressure bound is violated, the micro-kernel is using more registers than available and hence leads to register spilling. Similarly, when load-to-fma latency ratio bound is violated, some of the loads are not hidden due to pipelining. The additional loads incur a cost of loading from the higher level cache.

\begin{figure}
    \includegraphics[width=\linewidth]{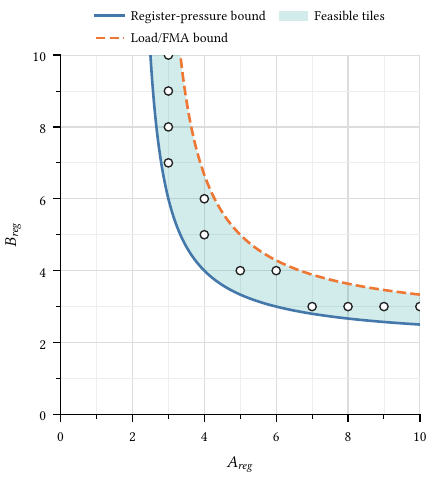}
    \caption{Candidate register tile sizes $(M_r, N_r)$ for AMD `Zen 5' architecture. The candidates are generated by enumerating all possible values of $M_r$ and $N_r$ that satisfy the register pressure and load-to-fma latency ratio bounds.}
    \label{fig:zen5-candidates}
\end{figure}

The equations, along with the assumption that $N_{reg}$ is small (typically $\leq 32$), effectively reduce the search space for optimal $M_r$ and $N_r$
(Figure~\ref{fig:zen5-candidates} highlights the possible candidates for `Zen 5').
Therefore, we are able to enumerate all possibilities and find the optimal one. To further narrow the list, we order them based on the number of bytes loaded from memory in one iteration. The candidate that loads fewer bytes is preferred since it minimizes the number of total memory accesses.

\subsubsection{Macro-kernel cost}

The micro-kernel cost described above accounts for all the compute cost of the GEMM operation. In contrast, the macro-kernel cost we describe below accounts for the cost of moving data between the primary memory and the L2 cache. The macro-kernel computes an output matrix of size ${M_c \times N_c}$ using inputs of size ${M_c \times K_c}$ and ${K_c \times N_c}$, and is where the bulk of data movement happens.

\begin{algorithm}[t]
    \caption{Macro-kernel cost $L_{\omega}$}
    \label{alg:macro-kernel-cost}
    \begin{algorithmic}[1]
        \Require Working-set size $S$, L2 size $S_{L2}$, L3 size $S_{L3}$, thread count $T$ for a given hardware configuration $H$
        \Procedure{CapacityMissCost}{$H, T, S$}
        \If{$S \leq S_{L2}$}
        \State \Return $0$
        \EndIf
        \State $S_{L3\text{-}resident} \gets \min(S, S_{L3}) - S_{L2}$
        \State $L \gets \Call{MoveFromL3Latency}{H, S_{L3\text{-}resident}}$
        \If{$S > S_{L3}$}
        \State $L \gets L + \Call{FetchFromMemLatency}{H, S - S_{L3}, T}$
        \EndIf
        \State \Return $L$
        \EndProcedure
    \end{algorithmic}
\end{algorithm}

When computing the output, the macro-kernel loop streams through either the $A$ or $B$ matrix. The number of cache misses depends on which matrix is streamed. Without loss of generality, we describe the macro-kernel cost assuming $B$ is streamed. We define the macro-kernel cost $L_{\omega}$ as,
\begin{equation}
    \begin{split}
        L_{\omega} = \Call{CapacityMissCost}{H, T, S_{A_c} +  S_{C_c}} \\
        + \; \beta \cdot \Call{CapacityMissCost}{H, T, S_{B_c}}
    \end{split}
    \label{eq:macro-cost}
\end{equation}

where $S_{A_c} = M_c * K_c * W_d$, $S_{B_c} = K_c * N_c * W_d$, and $S_{C_c} = M_c * N_c * W_d$ are the sizes of the input matrices A and B and output matrix C respectively. $\Call{CapacityMissCost}{H, T, S}$ denotes the cost of a capacity miss for a working set of size $S$ and is defined in Algorithm~\ref{alg:macro-kernel-cost}. Essentially, if the data set overflows L2, then we compute the cost of moving data from L3 to L2 and similarly if the data overflows L3, we compute the cost of fetching the overflowed data from memory to L3. We compute the capacity cost independently for the B matrix, and the A and C matrices together due to their differences in access patterns.
Additionally, when $K_c < K$, we account for the cost of accumulating the partial results by including the cost of loading the result matrix.

The parameter $\beta \in \left[0,1\right]$ in~\eqref{eq:macro-cost} is an architecture-dependent cost determined by the aggressiveness of the prefetcher. The prefetcher performs well when the data access pattern follows a fixed stride length, as it does when we stream through the B matrix in the macro-tile. Therefore, only a fraction of the B matrix incurs a capacity-miss cost. However, the A and C matrices are not streamed, so we do not expect the prefetcher to perform well.

\subsection{Packing}
\label{sec:packing}

GEMM libraries expect that the input matrices are pre-packed into memory a specific layout that depends on the tiling parameters. They expose routines to pack the matrices into the required format before performing the matrix multiplication. Without packing the inputs, the performance of the resulting GEMM operation degrades. The instruction used to load data in the innermost kernel differs from the optimal one, leading to inefficient memory accesses and suboptimal utilization of the cache hierarchy. Since the packing format depends on the tiling parameters, adopting a shape-dependent tiling strategy incurs additional costs.

However, with \codenamerefined{}, we identify a common packing strategy that works across different tiling parameters. Specifically, the packed layout should ensure two invariants: each access loads an entire cache line, and successive access locations follow a fixed stride length. We find that additional packing optimizations do not yield significant performance improvements and only lead to higher packing costs. Therefore, in \libraryname{}, we pack the input matrices into blocks of $(M_p \times K_p)$, where $M_p$ and $K_p$ are the least common multiples of all possible values of $M_r$ and $K_r$, respectively.

\section{Model based GEMM library}
\label{sec:library}

To demonstrate that the parameters obtained from \codename{} are realizable and achieve the expected performance, we implement a GEMM library \libraryname{} that uses the parameters derived from \codenamerefined{} for a given GEMM operation and hardware.

\libraryname{} is implemented in C++. In the library, we first use the tiling parameters generated by \codename{} to break down the large GEMM operation and then dispatch the micro-kernel to a JIT based small GEMM library included in \libraryname{}. To break down the large GEMM, we use \textsc{Parlooper}\cite{parlooper}, a library that dynamically generates loops with support for OMP based parallelization. Then, in the inner most loop, the JIT based small GEMM library generates the assembly instructions to compute the micro-kernel and caches it for subsequent uses. As an alternative to the JIT based GEMM library, we evaluated \textsc{Libxsmm}~\cite{libxsmm} and found it lacking the flexibility needed to generate kernels based on analytical-model-derived parameters. Therefore, we use our own JIT based small GEMM library.
In Section~\ref{sec:eval} we report the GEMM performance comparison with state-of-the art GEMM libraries and show that \libraryname{} is competitive.

\begin{figure}[t]
  \centering
  \resizebox{\linewidth}{!}{%
    \begin{tikzpicture}[
        font=\sffamily\scriptsize,
        >=stealth,
        flowbox/.style={
            draw=#1!75!black,
            fill=#1!8,
            rounded corners=2pt,
            line width=0.6pt,
            align=center,
            text width=1.85cm,
            minimum height=0.72cm,
            inner sep=3pt
          },
        stage/.style={
            draw=#1!80!black,
            fill=#1!10,
            rounded corners=2pt,
            line width=0.75pt,
            align=center,
            text width=1.95cm,
            minimum height=0.72cm,
            inner sep=3pt
          },
        arrow/.style={->, line width=0.8pt, draw=black!70},
        looparrow/.style={->, line width=0.8pt, draw=black!55, dashed},
        note/.style={font=\sffamily\tiny, text=black!72, align=center}
      ]

      \node[flowbox=blue] (seeds) at (-2.35, 0)
      {\textbf{Initial seeds}\\handwritten heuristic families};
      \node[flowbox=blue] (llm) at (-2.35, -1.35)
      {\textbf{LLM + config}\\prompt, models, MAP-Elites};
      \node[flowbox=blue] (cases) at (-2.35, -4.75)
      {\textbf{Pinned reference set}\\396 GEMM shapes};

      \node[stage=teal] (generate) at (0, 0)
      {\textbf{Seed programs}\\candidate C++ heuristics};
      \node[stage=teal] (evolve) at (0, -1.35)
      {\textbf{OpenEvolve search}\\mutate, select, archive};

      \node[stage=orange] (splice) at (0, -2.55)
      {\textbf{Materialize candidate}};
      \node[stage=orange] (compile) at (0, -3.65)
      {\textbf{Compile + run}};
      \node[stage=orange] (score) at (0, -4.75)
      {\textbf{Score candidate}};

      \node[flowbox=green] (best) at (0, -6.05)
      {\textbf{Best heuristic}\\single-answer C++ decision function};

      \draw[arrow] (seeds.east) -- (generate.west);
      \draw[arrow] (llm.east) -- (evolve.west);
      \draw[arrow] (generate.south) -- (evolve.north);
      \draw[arrow] (evolve.south) -- (splice.north);
      \draw[arrow] (splice.south) -- (compile.north);
      \draw[arrow] (compile.south) -- (score.north);
      \draw[arrow] (cases.east) -- (score.west);
      \draw[looparrow] (score.east) -- ++(0.55,0) |- (evolve.east);
      \draw[arrow] (score.south) -- (best.north);

      \node[draw=blue!75!black, fill=blue!8, rounded corners=1pt,
        minimum width=0.26cm, minimum height=0.16cm, inner sep=0pt] (leginput) at (1.65, -5.72) {};
      \node[anchor=west] at (1.85, -5.72) {Inputs};
      \node[draw=teal!80!black, fill=teal!10, rounded corners=1pt,
        minimum width=0.26cm, minimum height=0.16cm, inner sep=0pt] (legsearch) at (1.65, -5.94) {};
      \node[anchor=west] at (1.85, -5.94) {Search};
      \node[draw=orange!80!black, fill=orange!10, rounded corners=1pt,
        minimum width=0.26cm, minimum height=0.16cm, inner sep=0pt] (legeval) at (1.65, -6.16) {};
      \node[anchor=west] at (1.85, -6.16) {Evaluation};
      \node[draw=green!75!black, fill=green!8, rounded corners=1pt,
        minimum width=0.26cm, minimum height=0.16cm, inner sep=0pt] (legout) at (1.65, -6.38) {};
      \node[anchor=west] at (1.85, -6.38) {Outcome};

    \end{tikzpicture}%
  }
  \caption{High-level workflow for evolving a Waypoint CPU split-factor heuristic. Seed heuristics and LLM-generated mutations are evaluated against pinned exhaustive-search optima; the resulting metrics drive selection of the best heuristic.}
  \label{fig:origami-evolution-flow}
\end{figure}
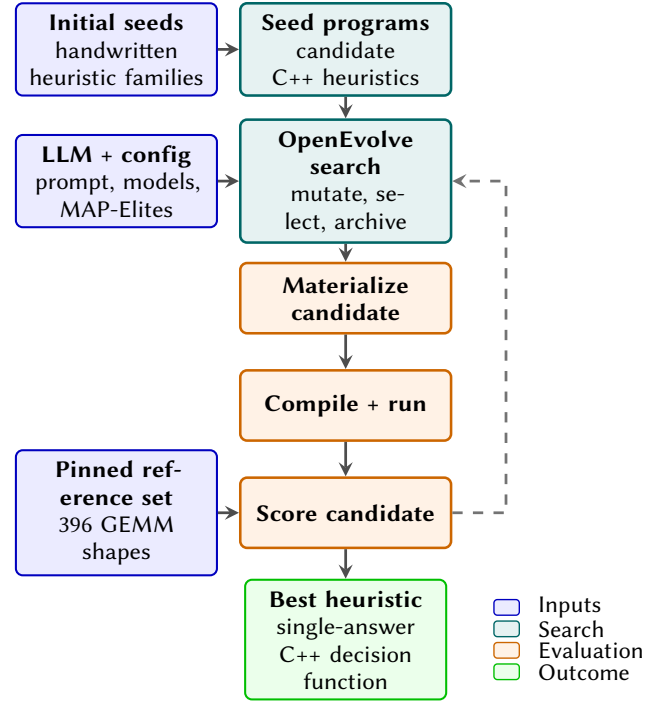

\libraryname{} supports two methods for finding the best tiling parameters: exhaustive search and fast heuristic-driven search. During exhaustive search, we evaluate all possible tiling parameters after pruning the search space based on insights from \codenamerefined{}. Specifically, we search only optimal $M_r$ and $N_r$ values and bound the $M_c$, $N_c$, and $K_c$ values based on the cache sizes.
However, for latency-sensitive applications such as LLM inference, the fast heuristic search offers a quick alternative.

The fast-heuristic search is developed using recent advances in algorithmic evolution \cite{barbariansatthegate,letbarbariansin,alphaevolve}. We use OpenEvolve \cite{openevolve}, an open-source implementation of AlphaEvolve \cite{alphaevolve}, to discover the heuristics. Our evolution process is focused on obtaining fast heuristics for a curated set of GEMM operations (or shapes) that modern LLM inference workloads invoke. However, the methodology can be used to discover heuristics that are tailored for any set of GEMM operations that the workload demands. The fast-heuristic generation is a one-time cost and the resulting heuristics can be used for all future GEMM operations.

Figure~\ref{fig:origami-evolution-flow} demonstrates the flow to obtain the heuristic search algorithm.
OpenEvolve starts with a set of seed programs as candidates for heuristics search. Our seeds are derived using insights from the tiling parameters generated by \codenamerefined{}. Then, OpenEvolve runs for many iterations, generating candidate heuristics in a genetic-mutation based algorithm. In each iteration we check the correctness of the tiling parameters and use
\codenamerefined{} is used to score the candidates. A candidate is good if the estimate of the selected tile parameter is not far from the best known. After running for ~100 iterations, OpenEvolve generates a fast-heuristic search algorithm that can be used to very quickly select near-optimal tiling parameters for a given set of GEMM operations and hardware. We defer the details of the OpenEvolve configuration to Appendix~\ref{sec:appendix_a}.

The resulting fast-heuristic search is $18851\times$ faster than the exhaustive search and discovers tiling parameters whose estimated latency is within $1.94\%$ (maximum) of the best tiling parameters derived by~\codename{}\footnotemark[1]. In our evaluation we show that the fast-heuristic search accelerates LLM performance on both current and future hardware. Our results include the cost of the fast heuristic search.

\section{Evaluation}
\label{sec:eval}

We evaluate the accuracy of our analytical model and the performance of \libraryname{} relative to three existing state-of-the-art libraries, \textsc{AOCL\textsubscript{TM}-BLAS}~\cite{amdblis}, \textsc{AOCL\textsubscript{TM}-DLP}~\cite{dlp}, and \textsc{SFC}~\cite{sfc}. \textsc{AOCL-BLAS} and \textsc{AOCL-DLP} are state-of-the-art tiling-based GEMM libraries that are optimized for AMD CPUs and support low-precision GEMM kernels used in machine learning workloads. While \textsc{AOCL-BLAS} uses handwritten micro-kernels, \textsc{AOCL-DLP} contains a JIT engine that dynamically generates the micro-kernel. In contrast, \textsc{SFC} is applicable to all x86 processors and implements a novel traversal mechanism (using space-filling curves) to minimize data access costs.

\begin{table}
    \caption{Evaluated GEMM shapes}
    \label{tbl:gemm_shapes}
    \begin{tabular}{c|c}
        \toprule
        Category                 & Count \\
        \midrule
        Small GEMM (<512)        & 4     \\
        Medium Square (512-4096) & 9     \\
        Large GEMM (>4096)       & 8     \\
        Tall skinny (M>>N)       & 4     \\
        Fat Short (M<<N)         & 2     \\
        Odd Shapes               & 4     \\
        LLM Shapes               & 20    \\
        \bottomrule
    \end{tabular}
\end{table}

Our evaluation is focused on the AMD Zen architecture. Specifically, we evaluate on `Zen 5' (Turin)~\cite{zen5} and `Zen 6' (Venice). Across both architectures, all our evaluations are performed for 3 different data types, FP32, BF16, and INT8. We evaluate the performance of the \libraryname{} for a set of 51 unique GEMM shapes (summarized in Table~\ref{tbl:evaluation_shapes}) (306 comparisons in total) that are commonly used in modern LLM inference workloads. A detailed list of the shapes used to evaluate can be found in Appendix~\ref{sec:appendix_b}.

With our evaluation we aim to answer the following research questions:
\begin{enumerate}
    \item[\textbf{RQ1}] Is the model accurate for different data types and architectures?
    \item[\textbf{RQ2}] How well does~\libraryname{} compare to state-of-the-art implementations?
    \item[\textbf{RQ3}] Does using the analytical model based \libraryname{} improve the performance of end-to-end LLM inference?
\end{enumerate}

The \codenamesimple{} and \codenamerefined{} models are evaluated against the Roofline model and the actual performance of GEMM operations implemented in the three libraries. \codenamesimple{} is tighter than Roofline and predicts actual performance to within $9.6\%$. \codenamerefined{} is accurate both in ranking different tiling options and in estimating the performance of the best tiling option compared with the libraries (geometric mean of $1.2\%$).

\libraryname{} achieves performance comparable to that of the three libraries across all configurations. When including packing costs, \libraryname{} beats \textsc{AOCL-BLAS} by $16.7\%$, \textsc{AOCL-DLP} by $13.6\%$, and \textsc{SFC} by $21.0\%$. Further, when comparing pure GEMM computation, \libraryname{} is $10.8\%$ faster than \textsc{AOCL-BLAS} and achieves performance similar to \textsc{AOCL-DLP} ($0.85\%$ slower) and \textsc{SFC} ($0.8\%$ faster).

Finally, we evaluate end-to-end LLM inference performance using PACE\textsubscript{TM}~\cite{amd_pace_2025}, a vLLM-compatible framework with optimized inference kernels for AMD CPUs. We embed the fast-heuristic search in PACE and evaluate it against a baseline implementation that does not use~\codename{}. We find that the shape-adaptive tiling strategy of~\codename{} yields average time-to-first-token (TTFT) speedups of $1.36\times$ to $1.98\times$ across different batch sizes, input lengths, models, and hardware architectures.

For the remainder of the section, we first describe the experimental setup and then elaborate on the results for each of the research questions.

\subsection{Experimental Setup}

The models are compared against two hardware architectures. One, AMD `Zen 5' (Turin) EPYC 9755 128-Core Processor with 512MB of L3 cache running at a fixed frequency of 2.7GHz. The other, AMD `Zen 6' (Venice) EPYC 9996 256-Core Processor with 1GB of L3 cache running at a fixed frequency of 2.5GHz. We were provided advanced access to `Zen 6' processor by AMD. While production systems are run with variable frequency, we fix the frequency to ensure consistent performance measurements and to isolate the performance improvements of our approach. We present comparison numbers to \textsc{AOCL-BLAS} and \textsc{AOCL-DLP} for all 3 data types (Int8, BFloat16, and FP32), and \textsc{SFC} for only 2 data types (BFloat16 and FP32) since \textsc{SFC} does not support Int8. The \libraryname{} is implemented in C++ and compiled using GCC 13.3.0. All comparisons are done by measuring cycles using the \texttt{rdtsc} instruction. To mitigate the effects of randomness when measuring the performance of GEMM libraries, for each shape, we run 10 warm-up trials and 100 measurement trials and report the best latency numbers. All numbers are reported using geometric mean.

\begin{table}
    \caption{Roofline vs \codenamesimple{} comparison for different data types across architectures. We report the median, mean, and max ratios of \codenamesimple{} to Roofline estimates across different shapes. A value of `1.000x` means \codenamesimple{} and Roofline estimate match.}
    \label{tbl:roofline_simple}
    \begin{tabular}{c|c|c|c|c|c}
        \toprule
        Datatype & Threads & n  & Median & Mean   & Max    \\
        \midrule
        Int8     & 1       & 58 & 1.001x & 1.005x & 1.031x \\
        Int8     & 2-8     & 24 & 1.000x & 1.001x & 1.011x \\
        Int8     & >8      & 20 & 1.000x & 1.001x & 1.002x \\
        BFloat16 & 1       & 58 & 1.001x & 1.004x & 1.031x \\
        BFloat16 & 2-8     & 24 & 1.000x & 1.002x & 1.013x \\
        BFloat16 & >8      & 20 & 1.001x & 1.001x & 1.005x \\
        Float32  & 1       & 58 & 1.001x & 1.004x & 1.031x \\
        Float32  & 2-8     & 24 & 1.001x & 1.003x & 1.014x \\
        Float32  & >8      & 20 & 1.002x & 1.002x & 1.006x \\
        \bottomrule
    \end{tabular}
\end{table}

For measuring end-to-end LLM inference times, we integrate \libraryname{} into AMD-PACE\cite{amd_pace_2025}. Our measurement compares against a baseline version that uses TPP~\cite{tpp} to dispatch GEMM calls. Our inference runs on 128 threads.

\subsection{RQ1: Model accuracy}

Table~\ref{tbl:roofline_simple} shows the comparison of \codenamesimple{} against the Roofline model across different data types and architectures. As shown, \codenamesimple{} is tighter than the Roofline model. Specifically when the data needed for computation is large, \codenamesimple{} deviates from the Roofline (up to 1.031x) accounting for the storage, and multi-threaded execution costs.

\begin{figure}
    \caption{\codenamerefined{} vs \codenamesimple{} vs Implementations (\textsc{AOCL-BLAS}, \textsc{SFC}, and \textsc{AOCL-DLP}). We report the ratios and distribution for all the shapes.}
    \label{fig:refined_comparison}
    \includegraphics[width=\linewidth]{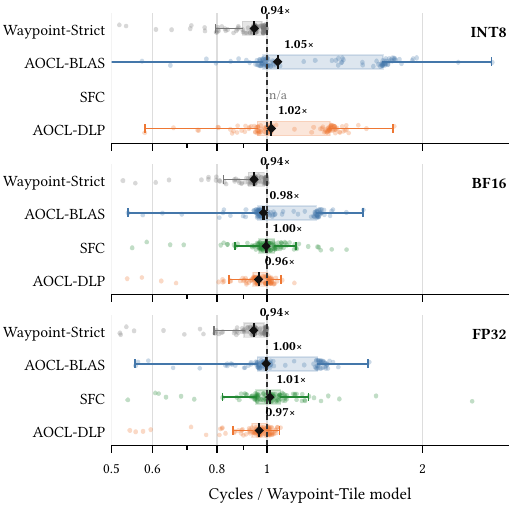}
    \Description{A comparison of the \codenamerefined{}, \codenamesimple{}, and actual implementations for various GEMM shapes.}
\end{figure}

\codenamerefined{} predicts the performance of a given tiling solution more accurately than \codenamesimple{} but does not guarantee a tight upper bound. Figure~\ref{fig:refined_comparison} compares the predictions of \codenamerefined{} and \codenamesimple{} with the measured performance of the three implementations. As shown, \codenamesimple{} serves as a tight upper bound on GEMM performance. The depicted \textsc{AOCL-BLAS} performance measurements form a bimodal distribution, with the two modes corresponding to the different architectures (`Zen 5' and `Zen 6').

Further, the \codenamerefined{} is able to accurately rank different tiling options for a given GEMM shape. To ensure the correctness of the ranking, we evaluate the 1st, 5th, and 10th tiling options for each GEMM shape and compare the predicted performance against the measured performance. We find that the \codenamerefined{} is able to accurately rank the top tiling option for
$88.7\%$ (within $5\%$ of the peak performance) of the evaluated GEMM shapes and strictly the best in $62.7\%$ of the cases.

\paragraph{Future hardware predictions}

\begin{table}
    \centering
    \small
    \setlength{\tabcolsep}{3pt}
    \caption{Predicted geometric-mean GEMM speedup from the proposed AI Compute Extensions (ACE) across 44 shapes per datatype vs `Zen 5' hardware. ACE throughput is effective number of Tile operations that can be issued per cycle.}
    \label{tbl:ace_predictions}
    \begin{tabular}{c|c|c}
        \toprule
        ACE throughput & BFloat16 & Int8   \\
        \midrule
        $1$            & 7.21x    & 7.45x  \\
        $2$            & 12.84x   & 13.48x \\
        $4$            & 21.47x   & 23.00x \\
        \bottomrule
    \end{tabular}
\end{table}

A key contribution of \codename{} is its ability to predict the performance of future proposed hardware. To demonstrate this, we evaluate the performance of the GEMM operation on a hypothetical future hardware that implements the proposed AI compute instructions~\cite{ace_spec} where one Tile operation computes a $16\times 16\times 4$ (for Int8) and $16\times 16\times 2$ (for BFloat16). Table~\ref{tbl:ace_predictions} shows the predicted performance gap of the GEMM averaged across all shapes for different throughput configurations of the AI compute instructions relative to `Zen 5' hardware. In `Zen 5', the VNNI instruction can be issued at the rate of 2/cycle and the tile op does the work equivalent to 16 VNNI instruction. As shown in the table, the model tempers the expected performance improvements of $8\times$ (1 tile op/cycle),$16\times$, and $32\times$, accounting for the data movement costs when compute is accelerated.
The result shows that adding many compute units will not automatically scale the performance linearly because the kernels will become memory bound.

\subsection{RQ2: \libraryname{} performance comparison}

\begin{figure*}
    \centering
    \includegraphics[width=\textwidth]{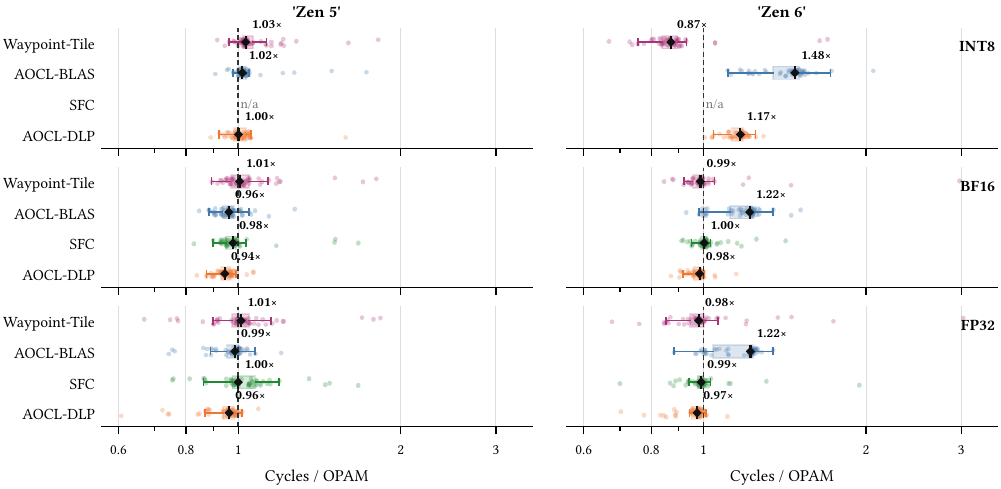}
    \caption{Relative performance of \textsc{AOCL-BLAS}, \textsc{SFC}, \textsc{AOCL-DLP}, and the \codenamerefined{} vs \libraryname{}, separated by hardware architecture.}
    \label{fig:opam-relative}
\end{figure*}

Figure~\ref{fig:opam-relative} compares the performance of the \libraryname{} relative to three state-of-the-art implementations and the \codenamerefined{} estimates for the same set of GEMM shapes. We observe that the \libraryname{} closely follows the performance predicted by the \codenamerefined{}, demonstrating the accuracy and practical utility of our approach. On `Zen 5', \libraryname{} meets the performance of existing, highly optimized GEMM libraries. However on `Zen 6' the libraries are yet to be tuned (especially for the Int 8 datatype) as the hardware is not readily available, but \libraryname{} demonstrates a significant performance advantage over the existing implementations.

\textbf{Packing cost differences}
The different implementations also differ in their packing strategies. As mentioned in~\ref{sec:packing}, the implementations extract peak performance by assuming that the data is pre-packed into contiguous memory regions in a specific format. For example, \textsc{SFC} requires that the data is laid out in the sequence of the space filling curve. This additional transformation increases the packing cost compared to more straightforward packing strategies used in \textsc{AOCL-BLAS}, \textsc{AOCL-DLP}, and \libraryname{}.

\begin{figure}
    \includegraphics[width=\linewidth]{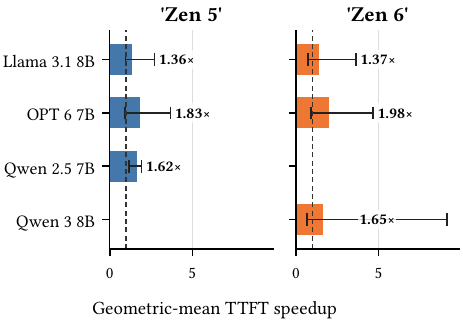}
    \caption{End-to-end inference speedup of \codename{} on prefill heavy LLM workloads across different hardware and models. Bars show the geometric mean; whiskers show the full range across batch sizes and input lengths.}
    \label{fig:pace-speedup}
\end{figure}

When comparing GEMM performance, \textsc{SFC} often achieves higher throughput due to its optimized data layout and tiling strategies. However, this comes at the cost of increased packing overhead. Comparing the packing costs, we find that \libraryname{} achieves a $1.40\times$ speedup over \textsc{AOCL-DLP}, $1.29\times$ over \textsc{AOCL-BLAS}, and $2.44\times$ over \textsc{SFC}.
We include a detailed table in Appendix~\ref{sec:appendix_b}. The higher packing cost can offset the performance gains in scenarios where the packing overhead is significant relative to the time taken to compute the result.

\textbf{Discussion} Including both the packing cost and GEMM latency, we observe that \libraryname{} is overall $1.168\times$ faster than \textsc{AOCL-BLAS}, $1.210\times$ faster than \textsc{SFC}, and $1.136\times$ faster than \textsc{AOCL-DLP}.
Conceptually, the gains of \libraryname{} over \textsc{AOCL-BLAS} and \textsc{AOCL-DLP} are primarily due to its shape-adaptive tiling strategies and comparable packing cost. In contrast, the gains over \textsc{SFC} are primarily driven by significantly lower packing cost and comparable GEMM performance matching the data reuse achieved by the optimized SFC algorithm.

\subsection{RQ3: End-to-end LLM inference performance}

We evaluate the impact of~\codename{} on end-to-end inference performance of Large Language models using, \textsc{AMD-PACE}~\cite{amd_pace_2025}.
Figure~\ref{fig:pace-speedup} shows the average end-to-end inference speedup on prefill heavy inference workloads that primarily rely on the performance of GEMM calls. Other workloads that involve more GEMV operations retain existing performance levels. As shown, \codename{} consistently improves inference performance across different hardware and model configurations.

The graph depicts speedups across batch sizes of $8,16,32,64$, and $128$ for input lengths of $1024$ and $2048$. With higher batch sizes, the cost of inference heavily skews towards GEMM operation latency. Therefore, by integrating \codename{}, we see the highest gains in higher batch sizes (e.g. $3.6\times$ for batch size $128$ on `Zen 5' with OPT 6 7B and $9.15\times$ for batch size $128$ on `Zen 6' with Qwen 3 8B). The gains are attributed primarily to improvements in GEMM computation times of the odd shapes encountered in LLM inference, driven by the shape-adaptive tiling strategy of~\codename{}. The graph also shows the ability of \codename{} to accelerate inference on new hardware. The range of speed up is larger in `Zen 6' compared to `Zen 5'.

\subsection{Extending to other architectures}

While our focus in the evaluation is on AMD x86 processor family, the principles used in both \codenamesimple{} and \codenamerefined{} are general and can be easily extended to other architectures. The estimates of performance depend primarily on a layered memory hierarchy and a pipelined vector/matrix processing unit.

The model immediately translates to Intel x86 processors given their similarity in ISA and hardware architecture to AMD x86 processors. Further, the model can be adopted to ARM processors. ARM processors also feature a layered memory hierarchy and pipelined vector/matrix processing units, although the specific details of the cache sizes, memory bandwidth, and vector instruction sets differ. Specifically, the variable vector length in ARM's Scalable Vector Instruction and Scalable Matrix Extension fit well with the parametric nature of \codenamesimple{} and \codenamerefined{}. The length and number of lanes of the registers are parametric in the model and can be configured based on the hardware architecture.

\section{Related work}
\label{sec:related}

There has been extensive work on modeling and optimizing the performance of General Matrix Multiplication (GEMM) operations on various hardware architectures.

\textsc{GotoBLAS}~\cite{goto-gemm} introduced a layered approach to implement GEMM operations in the BLAS standard library~\cite{blas}. They identified the optimization that most libraries~\cite{blis,amdblis,dlp} today adopt. Namely, that breaking down the large GEMM operation into smaller blocks (tiles) significantly improves cache utilization and overall performance. Subsequent works \cite{autotuning,atlas,opentuner} developed empirical methods to identify the optimal tiling parameters for a given hardware architecture and study the effects of different tiling choices on performance~\cite{autotvm}.

However, to overcome the time-consuming and resource intensive process of empirically identifying the best tiling parameters, analytical model approaches~\cite{analyticalmodellingisenough,model_after_atlas_1,model_after_atlas_2} have been proposed. These models~\cite{analytical_model_1} adapt to different hardware and to specific classes of shapes~\cite{analytical_model_2}. \codename{} extends prior works to identify optimal tiling parameters by being parametric and general not just in the hardware architecture but also in the specific input matrix shape.

The performance of tile-based GEMM implementations are sensitive to the data-layout of the input matrices in memory. To alleviate this, recent works have borrowed ideas from communication-avoiding algorithms~\cite{carma,sfc} to reduce data movement and improve cache utilization. They propose and show that by carefully organizing the input matrices in the order of a Space Filling Curve reduces cache misses and helps extract peak performance. \codename{} models are general enough to accurately predict the performance of such alternative implementations while at the same time, matching their performance with \libraryname{}.

While the traditional Roofline model~\cite{roofline} has been widely adopted to predict peak performance on proposed hardware designs, it fails to provide realistic estimates. As a result, the software never ceases to be optimized on existing hardware and hardware designers are unable to predict the realistic achievable performance on proposed hardware. Execution-Cache-Memory (ECM)~\cite{ecm,cacheawaremodel}, instruction-level models~\cite{instrmodel}, and GPU specific models~\cite{delta} have extended the Roofline model to capture different architectural effects and provide more realistic estimates. However, these models are either heavily tailored to a specific architecture, or require execution traces to be collected from the target hardware. \codename{} furthers the research by being both accurate and parametric in the hardware architecture, allowing hardware designers to get reliable estimates of the realistic performance achievable.

Our work is closely motivated by a similar effort focused on GPUs~\cite{tritonblas} which predicts performance of GEMM operations on GPUs by accounting for cache-miss rates, memory access overheads and other architectural effects. Our work \codename{}, generalizes the approach while demonstrating its applicability to CPUs.

\section{Discussion and Conclusion}
\label{sec:conclusion}

Under multi-threaded GEMM operations, \codename{} makes simplistic assumptions on the strategy to split work among different threads such as equal distribution of tiles or static scheduling, which may not always reflect the optimal parallel execution strategy. We observe that for certain input matrix sizes, using lower than the prescribed number of threads can yield better performance, highlighting the need for more sophisticated thread management strategies in future work.

We present \codename{}, a two part analytical model which first provides a tighter than Roofline estimate for GEMM performance and second, helps select the optimal tiling parameters to realize the estimates. The model derives several insights into estimating, and narrowing the search for optimal tiling parameters, ultimately enabling high-performance GEMM across diverse hardware architectures. We implement a GEMM library~\libraryname{} that uses the analytical model of \codename{} to guide its tiling decisions, and show that it achieves performance close to the theoretical estimates. \libraryname{} matches the performance of state-of-the-art GEMM libraries on existing hardware while showing considerable gains on future hardware architectures. Finally, we show that using \libraryname{} and \codename{} speeds up end-to-end CPU inference tasks significantly, demonstrating the practical impact of our approach.

\newpage
\bibliographystyle{ACM-Reference-Format}
\bibliography{main}

\appendix
\section{Appendix A: Heuristic Search Details}
\label{sec:appendix_a}

We run OpenEvolve to discover fast-search heuristics that offer a fast alternative to the exhaustive search of~\codename{}. To discover the heuristics, we first curate a set of GEMM operations (or shapes) that modern LLM inference workloads invoke. Then, we build a baseline of best tiling parameters for each shape using~\codename{}.

Starting with a predefined set of seed programs that contain initial heuristics, OpenEvolve runs many iterations, generating candidate heuristics that are evaluated against the baseline. Below we elaborate on the details of the seed programs and the evaluation process.

\subsection{Seed programs}

We generate 6 seed programs each containing independent set of heuristics that target a specific class of the shapes. The goal with the seeds is to provide a diverse set of possible heuristics that could be used to arrive at the final program and allow OpenEvolve to mutate, combine, and evolve them over many iterations to discover effective fast-search heuristics. Each seed program corresponds to a single island of evolution in OpenEvolve. The seeds are,

\begin{enumerate}
    \item Tiling based primarily on data type and ratio of matrix dimensions.
    \item Uses L1/L2 cache sizes to bound the tiling options.
    \item Targets optimal multi-threading splits based on the shape of the GEMM operation.
    \item Use arithmetic intensity to guide the inner most tiling decisions.
    \item Target specifically the AI Compute Instructions / AMX instructions.
    \item Initially baseline or the previous best program.
\end{enumerate}

\subsection{Fast heuristic search results}

We plot below the average difference in expected latency between the fast heuristic search algorithm and the exhaustive search algorithm.

\begin{table*}
    \centering
    \caption{Average difference in expected latency between the fast heuristic search algorithm and the exhaustive search algorithm across different shape categories and architectures.}
    \label{tbl:fast_heuristic_search_results}
    \begin{tabular}{l|rrrrrrr}
        \toprule
        Shape category & Z5 Float & Z5 BF16 & Z5 Int8 & Z6 Float & Z6 BF16 & Z6 Int8 \\
        \midrule
        Small Square   & 0.002\%  & 0.000\% & 0.443\% & 0.010\%  & 0.018\% & 0.000\% \\
        Medium Square  & 0.144\%  & 0.060\% & 0.059\% & 0.242\%  & 0.251\% & 0.426\% \\
        Large Square   & 0.089\%  & 0.059\% & 0.055\% & 0.234\%  & 0.295\% & 0.514\% \\
        Tall Skinny    & 0.692\%  & 0.145\% & 0.473\% & 0.111\%  & 0.093\% & 0.006\% \\
        Fat short      & 0.110\%  & 0.121\% & 0.127\% & 0.106\%  & 0.114\% & 0.209\% \\
        Odd Shapes     & 0.698\%  & 0.651\% & 0.838\% & 0.710\%  & 0.698\% & 1.590\% \\
        LLM Shapes     & 0.250\%  & 0.200\% & 0.235\% & 0.081\%  & 0.074\% & 0.292\% \\
        Overall        & 0.247\%  & 0.157\% & 0.253\% & 0.200\%  & 0.211\% & 0.427\% \\
        \bottomrule
    \end{tabular}
\end{table*}

\section{Appendix B: GEMM evaluation shapes}
\label{sec:appendix_b}

\subsection{Evaluation shapes}

\begin{table*}
    \centering
    \caption{GEMM evaluation shapes. $M$, $N$, and $K$ are the matrix dimensions, and $T$ is the number of threads.}
    \label{tbl:evaluation_shapes}
    \scriptsize
    \setlength{\tabcolsep}{5pt}
    \begin{tabular}{rrrr|rrrr}
        \toprule
        $M$  & $N$    & $K$   & $T$ & $M$   & $N$    & $K$   & $T$ \\
        \midrule
        64   & 64     & 64    & 1   & 4096  & 64     & 4096  & 1   \\
        64   & 2048   & 64    & 1   & 4096  & 128    & 4096  & 1   \\
        96   & 96     & 96    & 1   & 4096  & 4096   & 4096  & 1   \\
        128  & 32     & 2880  & 1   & 128   & 14336  & 4096  & 4   \\
        128  & 128    & 128   & 1   & 128   & 28672  & 4096  & 4   \\
        128  & 2880   & 2880  & 1   & 128   & 128256 & 4096  & 4   \\
        128  & 2880   & 4096  & 1   & 2048  & 2048   & 2048  & 4   \\
        128  & 2880   & 5760  & 1   & 4096  & 4096   & 4096  & 4   \\
        128  & 4096   & 128   & 1   & 8192  & 8192   & 8192  & 4   \\
        128  & 4096   & 4096  & 1   & 128   & 14336  & 4096  & 8   \\
        128  & 4096   & 14336 & 1   & 128   & 28672  & 4096  & 8   \\
        128  & 5120   & 2880  & 1   & 128   & 128256 & 4096  & 8   \\
        128  & 5760   & 2880  & 1   & 2048  & 2048   & 2048  & 8   \\
        128  & 6144   & 4096  & 1   & 4096  & 4096   & 4096  & 8   \\
        128  & 14336  & 4096  & 1   & 8192  & 8192   & 8192  & 8   \\
        128  & 28672  & 4096  & 1   & 128   & 14336  & 4096  & 16  \\
        128  & 128256 & 4096  & 1   & 128   & 128256 & 4096  & 16  \\
        256  & 256    & 256   & 1   & 2048  & 2048   & 2048  & 16  \\
        288  & 192    & 384   & 1   & 4096  & 4096   & 4096  & 16  \\
        512  & 512    & 512   & 1   & 8192  & 8192   & 8192  & 16  \\
        768  & 512    & 256   & 1   & 16384 & 16384  & 16384 & 16  \\
        992  & 768    & 1024  & 1   & 8192  & 8192   & 8192  & 32  \\
        1024 & 1024   & 1024  & 1   & 16384 & 16384  & 16384 & 32  \\
        2048 & 64     & 2048  & 1   & 16384 & 16384  & 16384 & 64  \\
        2048 & 128    & 2048  & 1   & 16384 & 16384  & 16384 & 128 \\
        2048 & 2048   & 2048  & 1   &       &        &       &     \\
        \bottomrule
    \end{tabular}
\end{table*}

Table~\ref{tbl:evaluation_shapes} lists the 51 unique GEMM shapes used in our
evaluation.  Each shape is evaluated on both hardware configurations and for
all three data types described in Section~\ref{sec:eval}.

\subsection{Packing cost}

Table~\ref{tbl:packing_cost} contains the packing cost speedup of \libraryname{}. Values are geometric-mean ratios of implementation packing cycles to \libraryname{} packing cycles for all the shapes.

\begin{table*}
    \centering
    \caption{Packing cost relative to \libraryname{}. Values are geometric-mean ratios of implementation packing cycles to \libraryname{} packing cycles for different thread counts.}
    \label{tbl:packing_cost}
    \begin{tabular}{p{2cm}|c|c|c|c}
        \toprule
        Hardware / Impl & Type     & $1$          & $2$--$8$     & $>8$          \\
        \midrule
        \multicolumn{5}{c}{\textbf{`Zen 5'}}                                     \\
        \midrule
        \multirow{3}{*}{\textsc{AOCL-BLAS}}
                        & INT8     & $2.02\times$ & $3.01\times$ & $10.34\times$ \\
                        & BFloat16 & $0.40\times$ & $1.37\times$ & $4.40\times$  \\
                        & Float32  & $1.11\times$ & $1.78\times$ & $5.05\times$  \\
        \midrule
        \multirow{3}{*}{\textsc{AOCL-DLP}}
                        & INT8     & $2.11\times$ & $3.75\times$ & $10.23\times$ \\
                        & BFloat16 & $0.44\times$ & $1.42\times$ & $4.48\times$  \\
                        & Float32  & $1.17\times$ & $1.89\times$ & $5.05\times$  \\
        \midrule
        \multirow{2}{*}{\textsc{SFC}}
                        & BFloat16 & $2.05\times$ & $1.64\times$ & $1.73\times$  \\
                        & Float32  & $3.92\times$ & $2.26\times$ & $3.66\times$  \\
        \midrule
        \multicolumn{5}{c}{\textbf{`Zen 6'}}                                     \\
        \midrule
        \multirow{3}{*}{\textsc{AOCL-BLAS}}
                        & INT8     & $1.62\times$ & $1.51\times$ & $1.94\times$  \\
                        & BFloat16 & $0.50\times$ & $0.65\times$ & $0.97\times$  \\
                        & Float32  & $1.01\times$ & $0.79\times$ & $1.52\times$  \\
        \midrule
        \multirow{3}{*}{\textsc{AOCL-DLP}}
                        & INT8     & $1.60\times$ & $2.93\times$ & $2.35\times$  \\
                        & BFloat16 & $0.55\times$ & $0.61\times$ & $0.98\times$  \\
                        & Float32  & $1.13\times$ & $0.86\times$ & $1.57\times$  \\
        \midrule
        \multirow{2}{*}{\textsc{SFC}}
                        & BFloat16 & $2.45\times$ & $1.15\times$ & $1.36\times$  \\
                        & Float32  & $3.49\times$ & $1.89\times$ & $3.55\times$  \\
        \bottomrule
    \end{tabular}
\end{table*}

\section{Model Assumptions}
\label{sec:appendix_c}

The estimates derived using \codename{} (both the Simple and \codenamerefined{}s) rely on certain assumptions about the behavior of the underlying hardware. We list below the assumptions along with the reasoning.

\textbf{Fixed frequency workloads:} \codename{} estimates the performance of a GEMM operation assuming that the CPU frequency is fixed. We calculate the number of cycles required to access data from memory using the frequency of the machine. Under a variable frequency scenario, one can extend the model by incorporating the dynamic changes in frequency to adjust the cycle estimates accordingly.

\textbf{Best instruction for compute:} Due to the highly optimized nature of GEMM kernels, \codename{} assumes that the best instruction i.e. the one with the highest throughput and lowest latency, is used for the compute operations. For example, in x86-ISA based CPU, \codename{} assumes that the AVX512 FMA/VNNI instruction is used compute operations. We do not explore the possibility of a mixed instruction use for a specific workload.

\textbf{Strided Data access:} \codename{} assumes that the data needed for compute is loaded into the registers in strides of contiguous memory locations. \codename{} accounts for the strided access by assuming the lowest cost for loading data into registers. For GEMM workloads, the assumption is valid since data is pre-packed to align accesses with cache lines and stride lengths.

\textbf{Cache memory model:} \codename{} assumes that the different level of caches provide different access latencies and that lower levels of caches have faster access times. We do not make any assumptions on the inclusivity, write-back policies, or replacement strategies of the caches. The model only considers the access latency differences between cache levels. A more complex model of the caches will help in determining the access latencies better.

\textbf{Partial bandwidth when loading data:} For multi-core (thread) GEMM operations, \codename{} assumes that the memory bandwidth is shared across all threads. Hence, each thread only receives a fraction of the total bandwidth available. The fractional cost is accounted for when computing the data accesses from memory.

\textbf{Peak store throughput:} We assume that stores occur at the peak throughput of the memory subsystem. Since the result-matrix locations targeted by store instructions typically reside in caches rather than main memory, the store operations are not expected to be bottlenecked by main memory and can achieve peak throughput.

\end{document}